\documentclass{amsproc}
\usepackage{amsfonts}

\usepackage{graphicx}

\theoremstyle{plain}

\numberwithin{equation}{section}

\input{tcilatex}

\begin{document}
\title{Local Quantum Theory beyond Quantization}
\author{Bert Schroer}
\address{Institut f\"{u}r Theoretische Physik\\
FU-Berlin, Arnimallee 14, 14195 Berlin, Germany\\
presently: CBPF, Rua Dr. Xavier Sigaud, 22290-180 Rio de Janeiro, Brazil }
\email{schroer@cbpf.br.}
\date{December 26, 1999}
\keywords{Quantum Field Theory, S-Matrix, Unruh Effect, Tomita Modular Theory }

\begin{abstract}
Recent progress on a constructive approach to QFT which is based on modular
theory is reviewed and compared with the standard quantization approaches.

Talk given at ``Quantum Theory and Symmetries'' Goslar, Germany, July 1999
\end{abstract}

\maketitle

\section{Local Quantum Physics beyond Quantization}

The slow but steadily increasing impact of the Tomita-Takesaki modular
theory on QFT over the last 20 years has presently reached a stage of
maturity where it promises to reshape the conceptual basis of local quantum
physics\cite{Bo}. Originally linked to Kubo-Martin-Schwinger (KMS) thermal
properties which characterize quantum statistical systems in the
thermodynamic limit \cite{H-H-W}, it became soon an important mathematical
concept in the pivotal\footnote{%
Different from QM which has to rely on an added interpretive setting, most
of its physical interpretation, thanks to causality of observables and
localization of states \cite{JPA}, is already contained within QFT.} issue
of localization \cite{B-W} and there exists by now an impressive body of
structural results obtained by modular methods, most of them have been
reviewed in \cite{Bo}.

The main goal in QFT is however not the structural results for their own
sake, but rather the classification of families of nontrivial field theories
and their constructive mathematical control; the structural theorems are
basically intuition- and confidence-creating intermediate steps in that
conquest. In this area we are in my view presently arriving at an important
change of paradigm in the development of QFT. The new message is that not
only are wedge-localized algebras important structural building blocks in
QFT (a fact which is not totally surprising in view of the importance of the
Rindler wedge in Unruh's discussion of the thermal Hawking-like aspect of
the light-front Horizon), but there are even concepts which allow to
classify and construct ``interacting'' wedge algebras as subalgebras of all
operators in the (incoming) Fock space of massive particles \cite{Sch1}.
Having a covariant net of wedge algebras (by acting with the
Poincar\'{e}-group on a standard wedge algebra), the nontriviality in the
setting of algebraic QFT just amounts to show that suitable intersections
(representing geometrically double cones) are not exhausted by multiples of
the identity. In particular one notices the existence problem approached in
this way is not directly threatened by short distance problems \cite{Sch2}.

Before explaining three basic new concepts which allow to pursue such a
program (and to analytically control it for nontrivial factorizing massive
models in two dimensions), it may be helpful to review the standard approach
and highlight some of its weak points. The QFT of almost all of the
textbooks uses a parallelism to classical field theory usually referred to
as ``quantization'' \footnote{%
This word covers a variety of different meanings ranging from the rigorous
functorial definition in connection with CCR and CAR algebras over classical
test function spaces for free fields, to the present more artistic use for
the implementation of interactions via Lagrangians. The use in the title of
the talk is the latter.}. This is most evident in the canonical approach in
the early days of QFT (Dirac, Heisenberg, Pauli and Fermi), but it also
remained visible in the renormalization theory of Tomonaga, Feynman,
Schwinger and Dyson as well as in the subsequent functional integral
approach based on euclidean actions. Even the so called causal approach of
Stueckelberg, Bogoliubov and Shirkov is not entirely independent of
classical ideas (although it goes a long way in this direction) because the
implementation of interactions by coupling free fields in a Fock space is
still in analogy to classical field theory.

Such quantization approaches usually involve some amount of ``artistry'' in
the sense that not all of the key requirements from which one starts are
reflected in the physical results. \textit{The physical} (renormalized) 
\textit{operators simply do not fulfill} (unless the model happens to be
superrenormalizable, which is certainly not the case for the more
interesting models) \textit{canonical commutation relations or functional
integral representations}. The only commutation structure which survives is
spacelike (anti)commutativity and Einstein causality. In QM the situation is
much better since e.g. path integral formulation has a solid mathematical
basis. But even there it is not advisable to use it outside the
quasiclassical approximation; it is extremely impractical to present a
course on QM with rich illustrations in such a setting. Artistry as
contained in the functional integral approach beyond QM is helpful as long
as the people who use it take it as a temporary recipe, and not for the
ultimate definition of what constitutes QFT.

The safest framework as far as avoiding such pitfalls is the so called
algebraic approach. It places the causality and localization structure into
the center of the stage and links it inexorably with the notions of
commutants in the theory of von Neumann algebras. It does not ban the good
old quantum fields completely, but attributes to them an auxiliary status
analogous to coordinates in differential geometry. One may use fields for
the generation of algebras, but there is nothing intrinsic or unavoidable
about them \cite{Ha}\cite{Bu-Ha}. It may be helpful to remind oneself that
the use of coordinates preceded the elegant intrinsic style of modern
differential geometry and even nowadays problem-adapted coordinates are
frequently used.

The problem with \ AQFT up to recent times was of course that in trying to
find a framework which avoids the above mentioned artistry\footnote{%
The prototype of a a balanced formulation of a quantum theory is of course
the operator/Hilbert space formulation of QM. AQFT tries to achieve such a
balance in the presence of relativistic causality/localization.}, it
suffered from an often criticized lack of practicality and constructiveness
even in such cases where its proponents used the word ``constructive''
instead of ``axiomatic'' in the titles of their papers. In the following I
would like to convince the reader that we are in the middle of a process of
change: the use of modular theory is rendering AQFT more computable and less
esoteric.

This turn of events was made possible because of three recent concepts which
went beyond the above mentioned result of Bisognano and Wichmann and which
are schematically explained in the sequel.

\begin{enumerate}
\item  It is possible to convert the unique Wigner one particle
representations $(m,s)$ directly into the net of free field algebras, thus
avoiding the intermediary use of field coordinatizations altogether \cite
{Sch1}\cite{BGL}. The construction uses the fact that the Wigner theory of
the positive energy representation of the connected part of the
Poincar\'{e}-group augmented by spacetime reflections on the rim of a wedge
allows to introduce an unbounded antilinear involutory pre-Tomita operator $s
$%
\begin{eqnarray}
s &=&kj\delta ^{\frac{1}{2}} \\
H_{R}(W) &=&real\,\,span\left\{ \phi \in H;s\phi =\phi \right\}  \notag
\end{eqnarray}
and to define a net of closed real subspaces $H_{R}(W)$ of the complex
Wigner representation space $H.$ In case of integer spin the requirement
that $k=1$ and $jH_{R}(W)=H_{R}(W^{opp})$ agrees with the geometric opposite
real localization subspace, whereas for semiinteger spin $k$ is a phase
factor related to the so-called Klein twist operator which preempts the
statistics on the level of the Wigner theory. For $d\geq 1+3$ the net of
real Hilbert spaces generated by Poincar\'{e} transformations is then
carried directly the net of wedge affiliated von Neumann algebras by the
Weyl (CCR) resp. the CAR functor and the Hawking-Unruh thermal aspect of
modular localization becomes manifest \cite{Sch1}.

In the case of d=1+2 and non semiinteger (anyone=nonquantized) spin, the
braid group statistics phase is already visible in the pre-Tomita theory in
Wigner space. In this case there exists no functor, the multiparticle space
does not have tensor product structure and the ``would be'' fields with
braid group statistics applied to the vacuum need the vacuum polarization
clouds to maintain the very braid group statistics. This net of localized
algebras cannot be obtained in the above functorial way (i.e. by rigorous
quantization) but has to be obtained by a method similar to that used in the
subsequent interacting case.

\item  For interacting bosonic/fermionic massive particles the Tomita
involution $J$ for the interacting wedge algebra has the following
representation in terms of the incoming free $J_{in}$%
\begin{equation}
J=S_{scat}J_{in}
\end{equation}
This relation can be derived from the TCP transformation of the S-matrix $%
S_{scat}$ and the close relation of $J$ to the TCP operator. In d=1+1
theories $J$ is equal to the TCP operator. The prerequisite is the validity
and completeness of the scattering theory which requires the standard mass
gap assumption (as in the LSZ scattering theory). The other modular object
of the pair ($\mathcal{A}(W),\Omega $) is the Lorentz boost \ This, as the
entire connected part of the Poincar\'{e} group, is independent of
interactions (at least in the bookkeeping based on scattering theory).

\item  The interacting wedge algebra has polarization free generators
(PFG's) $F(f)$ \cite{Sch1}\cite{Sch2} which create massive one particle
state vectors with a mass gap from the vacuum. 
\begin{equation}
F(f)\Omega =one-particle\,\,vector
\end{equation}
This property should be seen in the context of the following intrinsic
characterization of the presence of interactions (private communication by
D. Buchholz): operators associated to algebras localized in regions whose
causally completion is smaller than a wedge (i.e. the smallest region which
admits a L-boost automorphism) and which have a nonvanishing one-particle
component if acting on the vacuum create in addition a nontrivial
(interaction-determined) charge neutral polarization cloud (unless the net
is interaction free i.e. permits a free field generation). The existence of
PFG's is another illustration of the magic ability of QFT to maximally
utilize the breakdown of a physical argument (in this case a proof) for the
unfolding of a new unexpected phenomenon. The PFG-generators can be
explicitely charaterized in terms of their vacuum expectations. Although
they are determined in terms of the S-matrix (including the so-called
``bound state'' poles\footnote{%
Strictly speaking the hierarchy of elementary versus bound particle states
ceises to make sense in QFT where it must be replaced by the hierarchy of
charges. It has no intrinsic meaning to say e.g. that a particular particle
is a soliton independent of the chosen description (there are in general
various descriptions) unless the soliton charge explains all the other
charges by fusion and is not itself the result of more fundamental charge
fusion.}), they enter the vacuum expectation values in a novel ``nested''
fashion which gives rise to ``nondiagonal inclusive processes'' (referring
to the diagonal inclusive processes in cross sections) \cite{Sch2}. The
vacuum restricted to the wedge algebra generated by the PFG's is a thermal
KMS state of the type investigated by Unruh. For factorizing d=1+1 models
(examples: sine (sinh)-Gordon) the PFG's coalesce with a KMS representation
of the Zamolodchikov-Faddeev algebra thus equipping the latter for the first
time with a spacetime interpretation \cite{Sch1}\cite{Sch2}.
\end{enumerate}

This collection of obtained results requires some further comments. The net
of wedge algebras defined in terms of the previous PFG construction is not
sufficient in order to extract the physics. It is well known in algebraic
QFT that the full physical information is contained in the net of double
cone algebras and not yet in the wedge algebras. The calculation of the
former by intersections of wedge algebras is an unusual step, for which
presently efficient techniques are missing. It is the step from real
particle (on-shell) creation contained in the S-matrix to the virtual
polarization clouds which constitute the characteristic feature of QFT. It
is amazing that the wedge idea allows to divide the original problem into
two parts: the on-shell wedge algebra structure which is reminiscent of some
relativistic QM (with channel coupling between channels of different
particle numbers) whereas the off-shell double cone problem is QFT par
excellence. The wedge situation gives some justification to the (otherwise
quite meaningless) particle/field duality which sometimes is invoked in
analogy to the quantum mechanical particle/wave duality. The wedge
localization is the only one in which the two notions coexist: the particle
notion in the on-shell property of the wedge algebra and the field aspect in
the fact that the algebra has no annihilators and upon forming intersections
leads to the full off-shell virtual particle structure in a natural way
(i.e. without any further input). Everything is already preempted by the net
of wedge algebras: either the double cone algebras are void (only multiples
of the identity) in which case there is no local QFT compatible with the
wedge data, or the net of local observables exists and is unique.

It turns out that to decide which case appears in a given situation is quite
hard, even in the d=1+1 factorizing situation where the S-matrix commutes
with the incoming particle number. The point is that the intersection
condition which can be formulated as a vanishing commutator 
\begin{eqnarray}
&&\left[ A,F_{a}(f)\right] =0 \\
A &\in &\mathcal{A}(W),\,\,F_{a}(f)\in \mathcal{A}(W_{a})  \notag
\end{eqnarray}
where $F(f)$ are the wedge generators (see \cite{Sch2} for the notation) and 
$a$ denotes a spatial translation of $W$ into itself $W_{a}\subset W.$ The
coefficient functions of $A$ contained in the double cone relative commutant
in terms of the Z-F algebra generators can be computed and one finds the
so-called kinematical pole relation which describes the structure of the
multiparticle polarization clouds in terms of matrix elements of $A$ in the
``Zamolodchikov basis''. With other word the present formalism for the
calculation of the algebra of operators in the relative commutant $\mathcal{A%
}(W)^{\prime }\cap \mathcal{A}(W_{a})$ only determines them as bilinear
forms, which is not enough to calculate e.g. their correlation functions. To
specialists this problem is well known from the Karowski-Weisz-Smirnov
bootstrap-formfactor approach, where the control of convergence properties
in the construction of correlation function has (apart from some too simple
cases as the Ising order/disorder fields) remains as an unresolved issue. In
our formulation with nonpointlike finite localization regions the problem
looks somewhat simpler since (according to the Payley-Wiener theorem) the
momentum-space fall-off properties are better. There is however another
quite fundamental idea which one expects to lead to significant
simplifications. This is the idea to study instead of the wedge algebras for
massive d=1+1 models their \textit{chiral conformal holographic projections}%
. Intuitively this is related to light cone physics, but the standard
methods in this area are not good enough for the present purpose. The
conceptually as well as mathematically satisfactory method to handle such
holographic problems is the method of modular inclusions of subfactors \cite
{Wi}. These are subfactors (in the language of von Neumann algebras) inside
a larger factor whose modular group fulfills a certain consistency condition
with respect to the action of the larger modular automorphism on the
subfactor. The main theorem is that such situations are isomorphic to chiral
conformal theories in the sense of AQFT. It turns out that if one takes in
the above construction of relative commutants instead of a spacelike $a$ one
of the two lightlike translations $a_{\pm }$ one is precisely in that
situation of a modular inclusion. The associated chiral theory is localized
on the light cone. For the main theorem and its prove the reader is referred
to the literature \cite{Wi-S}\cite{GLRV}. It needs to be stressed that this
holographic association of a chiral conformal QFT to a massive theory is
quite different from its short distance universality class. Whereas the
fields of the latter generally decompose into tensor products belonging to
the two light rays, the holographic chiral theory remembers its origin
because it admits in addition to the 3 geometric Moebius transformations
another ``hidden'' (nongeometric) automorphism. This corresponds to the on $%
W $ locally acting opposite (-) light ray translation which if transferred
to the (+) ray becomes spread out i.e. totally ``fuzzy''. Here we used the
equality 
\begin{equation}
\mathcal{A}(\mathbb{R}_{+})=\mathcal{A}(W)
\end{equation}
between (half the) the light ray algebra and the wedge algebra has been made 
\cite{Sch2} as well as the fact that the natural nets for both algebras are
nonlocal relative to each other. The mass operator is neither given by $%
P_{+} $ nor by $P_{-}$ (which separately have a gapless spectrum), but
rather by the product of the geometric with the hidden generator $P_{+}P_{-}.
$

This kind of holography has a generalization to higher dimensions, but
instead of a $\left( d-1\right) $ dimensional light front theory (as one
would have naively expected), the formalism suggests to take $\left(
d-1\right) $ identical chiral theories carefully placed in a suitable
relative position in a common Hilbert space. A more appropriate terminology
would be to speak of a ``chiral scanning'' of a higher dimensional theory.
All this seems to be extremely powerful in a constructive approach, but it
has not yet been explored. The modular method applied to d=1+1 factorizing
systems in conjunction with its chiral holographic projection, with the aim
to obtain complete mathematical constructive control over this interesting
class of nontrivial 2-dimensional models is what I am presently working on 
\cite{Sch2}.

\section{Future Prospects}

The only way of giving some more hints in a page-limited conference report
is to mention some of the dreams which this new approach lends itself to.

\begin{itemize}
\item  \textbf{A purely field-theoretic classification and construction of
chiral theories based on modular theory}

The present constructions are using special (affine) algebras which did not
originate in QFT and which have no known higher dimensional generalizations.
On the other hand the exchange algebra which is of pure field theoretical
origin is (unlike its special case of CCR and CAR algebra) incomplete in
that its distribution theoretical character at colliding points remains
unspecified. With some hindsight and artistry on monodromy, one is able to
compute 4-point functions for the family of minimal models \cite{Re-S}\cite
{Ya}, but this is a far shot from. On the other hand a modular construction
would start as the exchange algebras from the statistics data (R-matrices
obtained by classifying Markov traces\footnote{%
This method (in the special case for which the braid group degenerates into
the permutation group $P_{\infty })$ appears already in the 70/71 DHR work 
\cite{Ha}} on the universal braid group $B_{\infty },\simeq $ topological
field theory) and use them (instead of the ) for the modular construction of
the halfline algebra. By Moebius transformation (instead of intersections)
one should be able to obtain the charge-carrying (which carry the plectonic
charges) operators for arbitrary small regions. In this way an old dream
74/75 of Swieca and myself could still come true which consisted in the hope
to classify the conformal block decompositions by purely field theoretic
methods (unlike the later use of representation theory of special (affine)
algebras) and in this way convert chiral QFT into a bona fide theoretical
laboratory of general QFT. This dream would also include the understanding
of the very peculiar and special Friedan-Qiu-Shenker quantization in the
energy-momentum tensor as a consequence of the\ more universal
(DHR-Jones-Wenzl) statistics quantization.

\item  \textbf{A theory of ``free plektons''}

It has been known for some time, that the d=1+2 carriers of braid group
statistics must have a semiinfinite spacelike string localization and that
they obey a spin-statistics theorem. Unlike chiral theories, 3-dim. theories
do allow for deformation parameters (coupling strength whose continuous
change have no effect on the superselection rules). It is an educated guess
that there exist ``free plektons'' in the sense that they have vanishing
cross sections although their S-matrix is not one but rather piecewise
constant and equal to one of the R-matrices (with all the caveats related to
the very definition of scattering matrices in the absence of a Fock space
tensor structure!). Such an S would still commute with the incoming particle
number (as in the d=1+1 factorizing case) and would be used in a modular
approach in order to construct PFG wedge algebra generators and, via
intersections, to obtain spacelike cone localized (string-like) plektonic
operators. Although without real particle creation, such a free plektonic
theory would (unlike free Bosons and Fermions) have a rich virtual vacuum
polarization structure \cite{Mu}. The nonrelativistic limit should be
carried out in such a way that the spin-statistics connection remains
intact. The nonrelativistic limit would maintain its field theoretic vacuum
polarization structure (necessary to upheld the spin-statistics relation)
and the result would therefore not be consistent with the well-known
Leinaas-Myrrheim or Aharonov-Bohm (in Wilczek`s approach) quantum mechanical
description. Plektons are also too noncommutative in order to permit a
well-defined Lagrangian quantization description.

\item  \textbf{Holographic or scanning methods for higher dimensional
constructions}

Difficult problems as the classification and constructive determination of
higher dimensional QFT`s often render themselves more susceptible to
solution if one succeeds to chop them into several easier pieces. The
``smallest'' and best understood kind of QFT is certainly chiral QFT.
According to previous comments, one may hope that a finite collection of
chiral theories in carefully arranged relative positions (modular
intersections and inclusions) in a common Hilbert space may contain all the
necessary information for the construction of a higher dimensional QFT.
\end{itemize}

Such a construction, if feasible, would sharpen a paradigm which underlies
AQFT\footnote{%
This has been formulated on different occasions in past articles and
conference reports by R. Haag}: the new physical reality (of AQFT or Local
Quantum Physics) is not that of a manifold with a material content (the
Newtonian reality which extends into relativity and QM), but is rather
thought of as coming about through relations between objects which
themselves have no individuality (like the monades of Leibnitz). In
technical-mathematical language the physical reality of local nets comes
about by modular inclusions/intersections (and more general kinds of
relative positioning) of the unique hyperfinite type III$_{1}$ von Neumann
factors which (apart from being able to be contained in each other and form
intersections) behave in many respects as points in geometry.

\end{document}